%% file: main.tex
\documentclass[]{spie}  

\usepackage{amsmath,amsfonts,amssymb}
\usepackage{graphicx}
\usepackage[colorlinks=true, allcolors=blue]{hyperref}
\usepackage{caption}
\usepackage{lineno}

\title{ComPair-2: A Next Generation Medium Energy Gamma-ray Telescope Prototype}

\author[a]{Regina Caputo}
\author[a]{Carolyn~Kierans}
\author[b,a,c]{Nicholas~Cannady}
\author[d]{Abe~Falcone}
\author[e]{Yasushi Fukazawa}
\author[f]{Manoj Jadhav}
\author[g]{Matthew Kerr}
\author[h]{Nicholas Kirschner}
\author[i,a,c]{Kavic Kumar}
\author[j]{Adrien Laviron}
\author[k]{Richard Leys}
\author[a,r]{Iker Liceaga-Indart}
\author[a]{Julie McEnery}
\author[f]{Jessica Metcalfe}
\author[i,a,c]{Zachary Metzler}
\author[l]{Nathan Miller}
\author[a]{John Mitchell}
\author[m]{Lucas Parker}
\author[k]{Ivan Peric}
\author[a]{Jeremy Perkins}
\author[g]{Bernard Phlips}
\author[a]{Judith Racusin}
\author[i,a,c]{Makoto Sasaki}
\author[n]{Kenneth N. Segal}
\author[g]{Daniel Shy}
\author[a,o]{Amanda L. Steinhebel}
\author[k]{Nicolas Striebig}
\author[e]{Yusuke Suda}
\author[p]{Hiroyasu Tajima}
\author[b,a,c]{Janeth Valverde}
\author[a,o]{Daniel P. Violette}
\author[g]{Richard Woolf}
\author[q]{Andreas Zoglauer}

\affil[a]{NASA Goddard Space Flight Center, Maryland, U.S.A}
\affil[b]{University of Maryland, Baltimore County, Maryland, U.S.A.}
\affil[c]{Center for Research and Exploration in Space Science and Technology, Maryland, U.S.A.}
\affil[d]{The Pennsylvania State University, Pennsylvania, U.S.A.}
\affil[e]{Hiroshima University, Hiroshima, Japan}
\affil[f]{Argonne National Laboratory, Illinois, U.S.A.}
\affil[g]{Naval Research Laboratory, Washington, D.C., U.S.A}
\affil[h]{The George Washington University, Washington, D.C., U.S.A}
\affil[i]{University of Maryland, College Park, Maryland, U.S.A.}
\affil[j]{Laboratoire Leprince-Ringuet, CNRS/IN2P3, {\'E}cole Polytechnique, Institut Polytechnique de Paris, Palaiseau, France}
\affil[k]{Karlsruher Institut f{\"u}r Technologie, Germany}
\affil[l]{John Hopkins University, Maryland, U.S.A}
\affil[m]{Los Alamos National Laboratory, New Mexico, U.S.A.}
\affil[n]{Heliospace Corporation, Berkeley, California, U.S.A.}
\affil[o]{NASA Postdoctoral Program Fellow}
\affil[p]{Nagoya University, Nagoya, Japan}
\affil[q]{Space Sciences Laboratory, University of California Berkeley, California, U.S.A.}
\affil[r]{Catholic University of America, Washington, D.C., U.S.A.}

\authorinfo{Further author information: (Send correspondence to R. Caputo and C. A. Kierans)\\R. Caputo: E-mail: regina.caputo@nasa.gov,   C. A. Kierans: E-mail: carolyn.a.kierans@nasa.gov}

\begin{document} 
\maketitle

\begin{abstract}
Many questions posed in the Astro2020 Decadal survey in both the New Messengers and New Physics and the Cosmic Ecosystems science themes require a gamma-ray mission with capabilities exceeding those of existing (e.g. \textit{Fermi}, \textit{Swift}) and planned (e.g. COSI) observatories. 
ComPair, the Compton Pair telescope, is a prototype of such a next-generation gamma-ray mission. It had its inaugural balloon flight from Ft. Sumner, New Mexico in August 2023. 
To continue the goals of the ComPair project to develop technologies that will enable a future gamma-ray mission, the next generation of ComPair (ComPair-2) will be upgraded to increase the sensitivity and low-energy transient capabilities of the instrument. 
These advancements are enabled by AstroPix, a silicon monolithic active pixel sensor, in the tracker and custom dual-gain silicon photomultipliers and front-end electronics in the calorimeter. 
This effort builds on design work for the All-sky Medium Energy Gamma-ray Observatory eXplorer (AMEGO-X) concept that was submitted the 2021 MIDEX Announcement of Opportunity. 
Here we describe the ComPair-2 prototype design and integration and testing plans to advance the readiness level of these novel technologies.  
\end{abstract}

\keywords{gamma-ray, future observatories, ComPair, AMEGO-X, AstroPix}

\section{INTRODUCTION}
\label{sec:intro}  
\input{Introduction}

\section{THE COMPAIR-2 INSTRUMENT}
\label{sec:instrument}
\input{Instrument}

\section{INTEGRATION AND TESTING}
\label{sec:IandT}
\input{IandT}

\section{PIPELINES, DATA ANALYSIS AND PERFORMANCE}
\label{sec:pipelines}
\input{Pipelines}

\section{CONCLUSIONS}
\label{sec:conclusions}
\input{Conclusions}

\acknowledgments       
 
This work is supported under NASA Astrophysics Research and Analysis (APRA) grants NNH17ZDA001N- APRA (improved Compton and pair event reconstruction), NNH18ZDA001N-APRA (AstroPix and dual-gain SiPM readouts for the CsI Calorimeter), and NNH22ZDA001N-APRA (ComPair-2).
The material is based upon work supported by NASA under award number 80GSFC21M0002. 

\newpage

\bibliography{report} 
\bibliographystyle{spiebib} 

\end{document}

%% file: Introduction.tex
The coordinated study of extreme explosions and accelerators is driven by astrophysical objects that emit photons and other messengers: gravitational waves, neutrinos, and cosmic rays. 
Gamma-ray observations are at the forefront of multimessenger discoveries, and the medium-energy gamma-ray band (100 keV to 100 MeV) plays a critical role in the identification and characterization of their sources.
This regime is not fully covered by current or future observatories 
and a future gamma-ray mission directly addresses needed capabilities to advance multimessenger astrophysics through spectroscopy, imaging, and polarization.
The ComPair-2 instrument serves as a prototype to such a next generation gamma-ray mission: the AMEGO-X MIDEX concept~\cite{AMEGOX}.

Over the past few years, the universe has revealed the necessity of gamma-ray observations for multimessenger astrophysics. 
In the summer 2017 when a binary neutron star system merged, an intense burst of gamma rays was the first detectable light from the cataclysm. 
Gravitational waves detected from this explosion set off a worldwide, multiwavelength campaign to observe GW170817~\cite{LIGOScientific:2017vwq}. 
A month after this discovery an active galactic nucleus (AGN), TXS 0506+056, was undergoing a bright gamma-ray flare and a high-energy neutrino detection by the IceCube observatory was observed to be spatially coincident~\cite{IceCube:2018dnn}.
This provided the first link between AGN and neutrinos and revealed the fundamental components of the accelerated particles. 
These watershed discoveries were enabled by NASA’s \textit{Fermi} Gamma-ray Space Telescope~\cite{lat_inst} and heralded a new era of multimessenger astrophysics.

AMEGO-X is a wide field-of-view, medium-energy gamma-ray telescope concept that would directly probe the physics at the heart of key multimessenger sources. 
The MeV band provides unique insight into the physics of binary neutron star mergers through the characterization of the associated short gamma-ray burst (GRB)~\cite{Burns:2019byj, Howell:2018nhu}.
AGN are theorized to be a major source of astrophysical neutrinos, and a coincident detection of photons in the MeV band would provide smoking-gun evidence of the hadronic processes in the jet~\cite{Lewis:2021roc, Keivani:2018rnh}. 
Coincident gamma-ray observations are needed to maximize the scientific return of the upcoming Cherenkov Telescope Array (CTA) observatory in the GeV/TeV gamma-ray regime, advanced LIGO/Virgo in gravitational waves, and IceCube-Gen 2 in neutrinos.
Furthermore, a telescope covering the 100~keV -- 1~GeV energy range complements the Compton Spectrometer and Imager (COSI) Small Explorer mission selected by NASA to fly in 2027~\cite{2021arXiv210910403T}. 
COSI will provide spectroscopic capabilities in the MeV band to image narrow emission lines from our Galaxy at energies below 5~MeV. 
AMEGO-X is a larger mission with a wider field-of-view, broader band pass, and unprecedented sensitivity required to enable priority time-domain and multimessenger science in the next decade.
More details about the AMEGO-X science case and instrument design can be found in Ref.~\citeonline{AMEGOX}.

%% file: Instrument.tex
\begin{figure}[ht]
    \centering
    \includegraphics[width=0.8\textwidth]{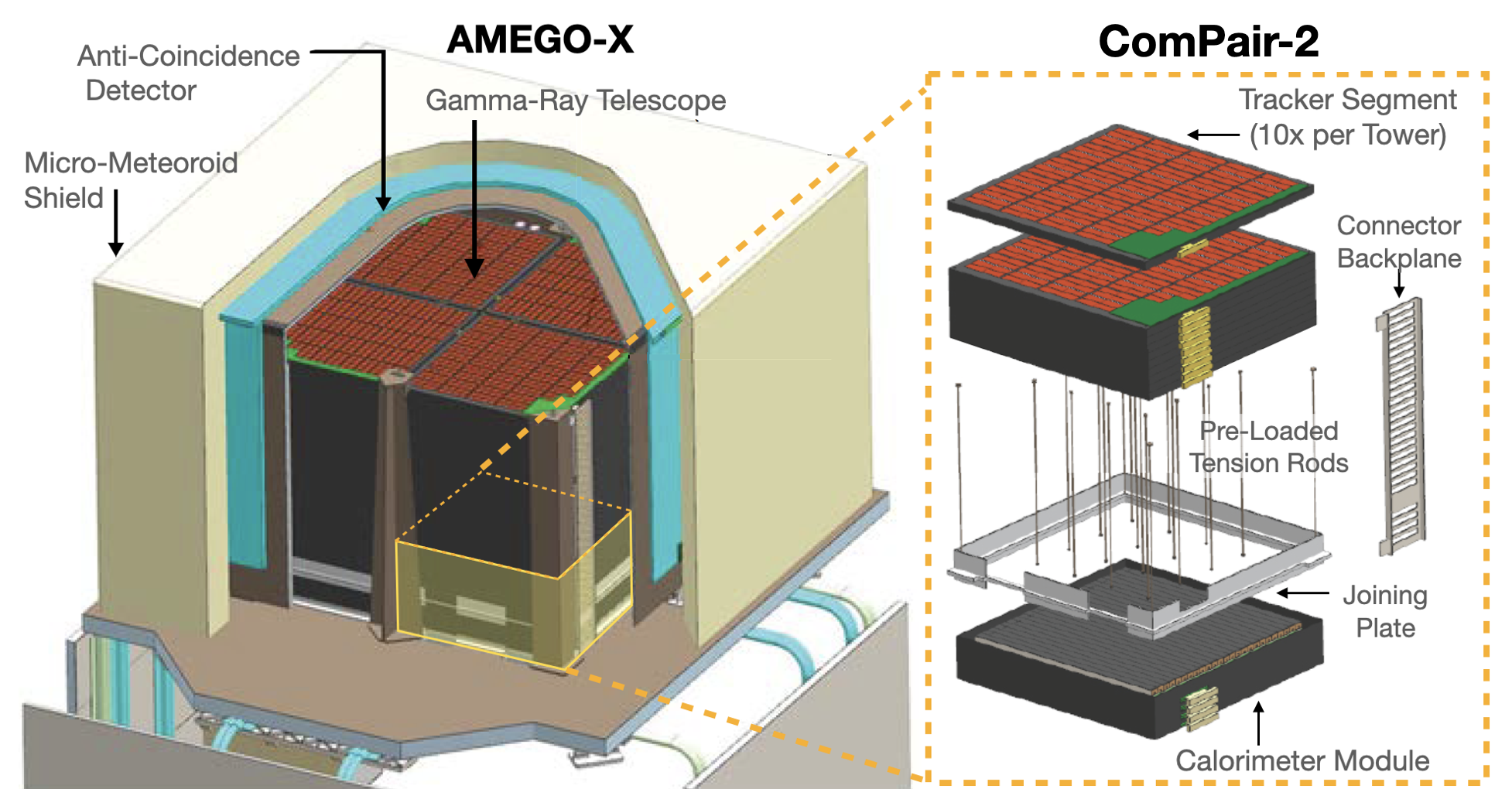}
    \vspace{0.3cm}
    \caption{\textit{The ComPair-2 instrument will advance and validate the the AMEGO-X hardware and software tools through a short, single tower, flight-like prototype. The build and environmental tests of ComPair-2
    will raise the Technical Readiness Level of the AMEGO-X Gamma-Ray Telescope demonstrating the feasibility of the telescope construction and design.}}
    \label{fig:compair_cad}
\end{figure}

The ComPair-2 instrument will advance and validate the AMEGO-X hardware and software tools by building and testing a flight-like prototype. The AMEGO-X Gamma-Ray Telescope, shown in \textbf{Figure~\ref{fig:compair_cad}}, consists of four identical towers, each with 40 Tracker Segments and a 4-layer Calorimeter module, all surrounded by an Anti-Coincidence Detector. The ComPair-2 instrument is a single, short tower comprised of a 10-layer Tracker and a full-scale Calorimeter. The integration and testing of ComPair-2, as well as the performance validation, will directly raise the Technology Readiness Level (TRL) of the AMEGO-X Gamma-Ray Telescope to 6 in preparation for a future MIDEX Announcement of Opportunity. 

\subsection{Concept of Operation}
\label{sec:conops}

\begin{figure}[h]
    \centering
    \includegraphics[width=0.47\textwidth,trim={0 5cm 0 0},clip]{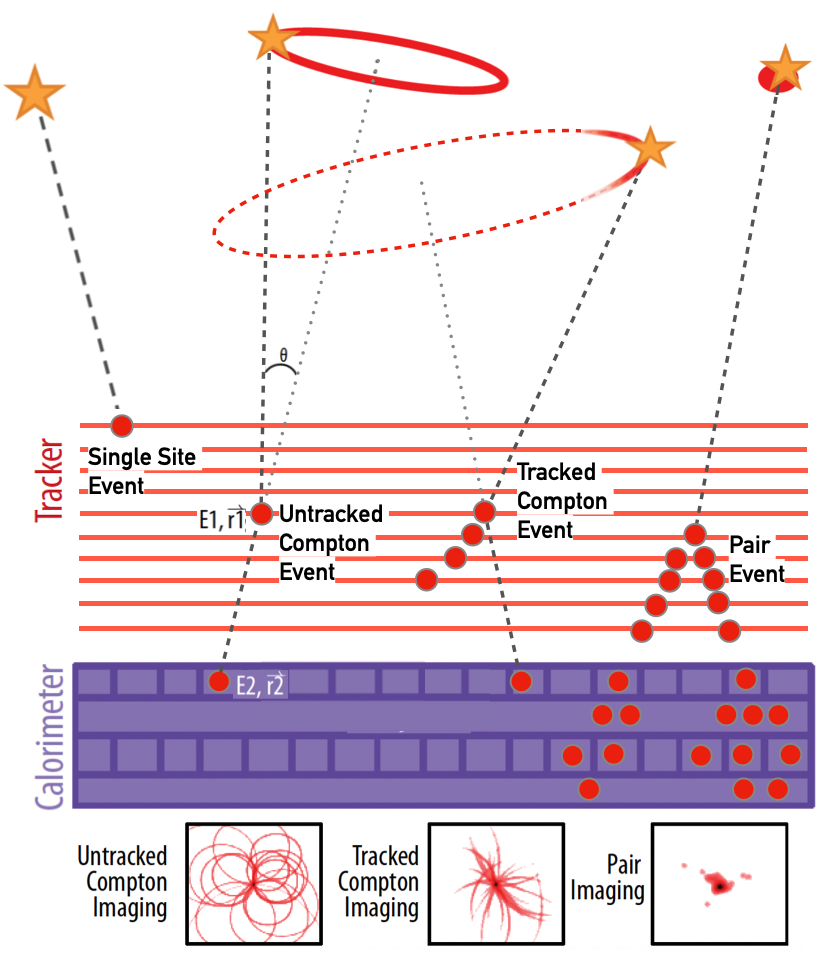}
    \caption{\textit{The ComPair-2 instrument detects and characterize photons across three orders of magnitude in energy. The interaction on the far left demonstrates the interaction of photons from 25 to $\sim$100 keV, where they will be absorbed in a single layer. Above 100 keV, events will Compton Scatter as illustrated by the two middle events. Above $\sim$10 MeV, photons will convert into electron and positron pairs leaving two tracks and a reconstructable vertex in the instrument, as shown in the far right. }}
    \label{fig:Compton_detection}
\end{figure}

The AMEGO-X instrument is optimized for continuum sensitivity across three orders of magnitude from 100~keV to 100~MeV, with a threshold of 25~keV for transient sources. 
This range covers three different photon-matter interactions, depending on the initial energy of the photon.  
ComPair-2 implements the same detection principles in a smaller detector volume. Four different event types and the resulting single-photon localization are shown for the ComPair-2 design in~\textbf{Figure~\ref{fig:Compton_detection}}.

\vspace{0.3cm}

Between about 100 keV and 10 MeV, photons predominantly Compton scatter. 
The measured position and energy of a Compton scattered interaction constrains the initial direction of the primary gamma ray to a circle in the sky~\cite{KieransCompton} (\textit{untracked Compton event}). 
Compton scattering is inherently polarization sensitive, and a linearly polarized source generates a sinusoidal azimuthal scattering angle distribution in the instrument~\cite{1997SSRv...82..309L}. 
If the direction of the first Compton-scattered electron is measured in the Tracker, this additional kinematic information constrains the photon direction to an arc and these \textit{tracked Compton events} allow for improved background rejection~\cite{AKYUZ2004127}.
Higher-energy gamma rays ($\gtrsim$10 MeV) convert to an electron-positron pair, which in turn is detected through ionization tracks in the Tracker.
For \textit{pair events}, the direction of the incoming photon is determined by the positions of the interactions in the Tracker and the total energy is determined by the electromagnetic shower(s) detected in the Calorimeter~\cite{lat_inst}.
At energies below the Compton regime ($\lesssim$100 keV), photons predominantly undergo photoelectric absorption in a single pixel, and for bright transients, such as GRBs, these interactions allow for higher detection rates, low-energy spectral information, and source localizations of a few degrees~\cite{2022ApJ...934...92M}.
The ComPair-2 Tracker enables short duration ($<$100 s) readout of \textit{single-site events} to measure emission of transient events down to 25 keV.

\subsection{AstroPix Tracker}
\label{sec:tracker}

The Tracker’s main functionality is to measure the energy and position of gamma-ray and charged-particle interactions with high precision, and it does so with pixelated silicon detectors. 
Following from ground-based particle physics heritage trackers~\cite{Peric:2007zz, Schoning:2020zed}, the AstroPix team has developed Active Pixel Sensors (APS) optimized for gamma-ray telescopes~\cite{BREWER2021165795,Steinhebel2023, suda2024},
where the major advancement is the reduction in noise enabled by the integrated readout electronics within the detecting material using proven high-voltage CMOS processing~\cite{peric_high-voltage_2018, peric_CMOS_aps_21}. 
Each pixel contains a charge-sensitive preamplifier and comparator, where the active circuitry results $<$1\% loss in charge-collection volume. 
The lower threshold results in higher effective area for Compton events and enables single-site event sensitivity down to 25 keV. 
The chips are 500~$\mu$m thick and will operate at full depletion with a bias voltage $\sim$300~V~\cite{steinhebel2024}.

\begin{figure}
    \centering
    \includegraphics[width=0.30\textwidth, clip=true]{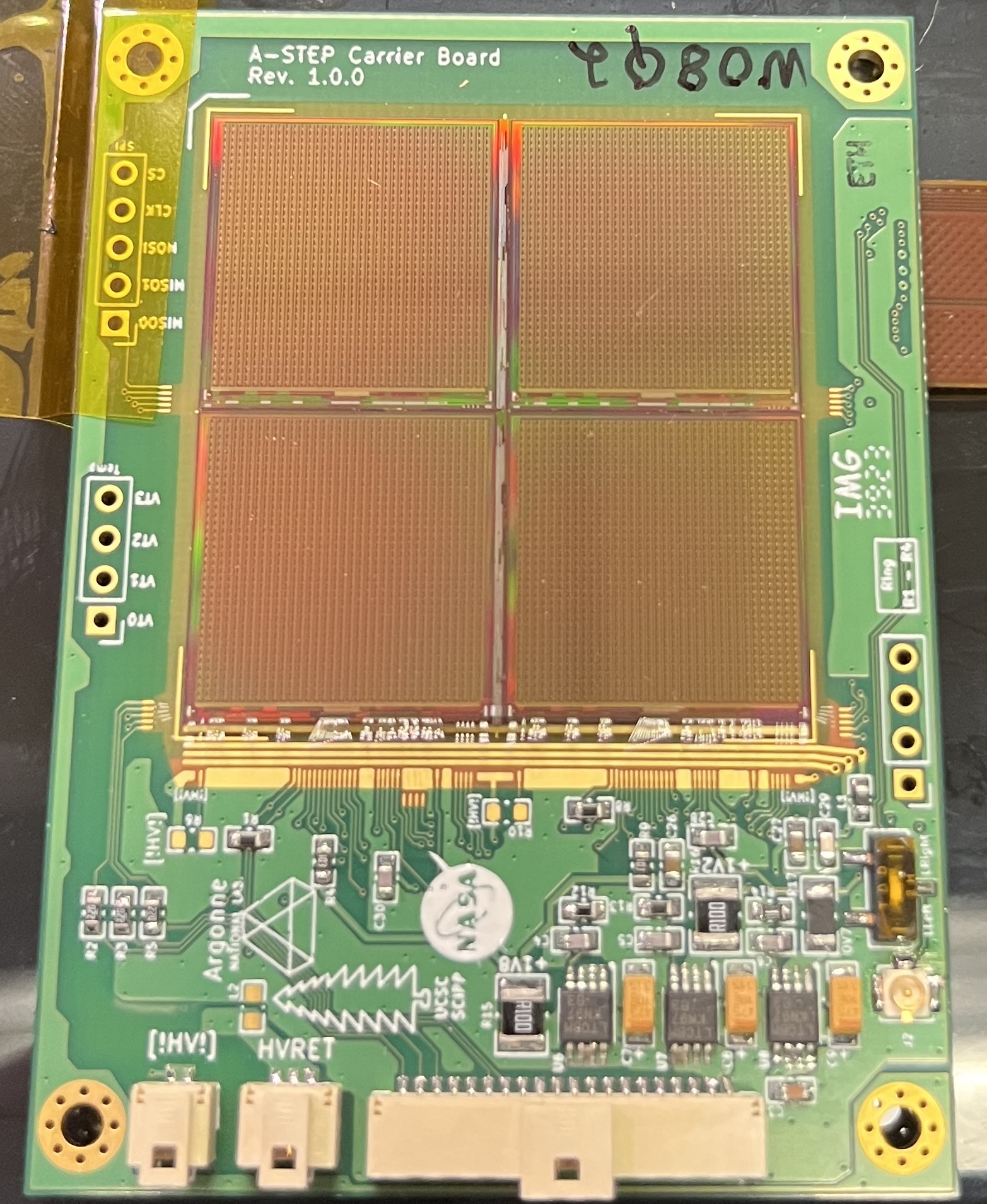}
    \includegraphics[width=0.68\textwidth, clip=true]{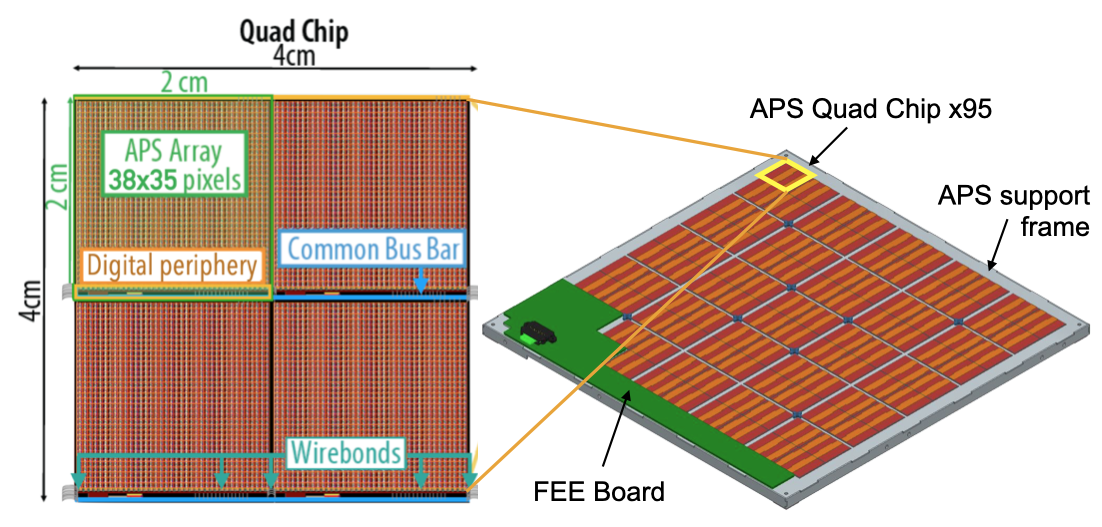}
    \vspace{0.3cm}
    \caption{\textit{The ComPair-2 Tracker uses AstroPix Quad Chips (left) leveraging CMOS technology to provide low energy response. A single exploded layer is shown on the right including the 95 Quad Chips (red), the carbon fiber frame (gray) and the Front End Electronics (FEE) board (green). A zoom-in of the AstroPix quad chip schematic is in the middle and the v3 Quad Chip mounted to a testing carrier board is shown on the left. }
    \label{fig:compair_tracker_layer}}
\end{figure}

The AstroPix signal digitization and readout circuitry is contained in a digital periphery on the bottom of each APS array.
The energy deposited in each pixel is determined by the time-over-threshold (TOT) of the amplified pulse with a 3.125~ns resolution.
The APS arrays use a daisy-chain Serial Peripheral Interface (SPI) to handle chip configuration and readout of the TOT data, event time stamp, and pixel address for each signal above threshold.
A flight-like version of the AstroPix detectors (v3) has been delivered and is currently being tested and characterized at GSFC, Argonne National Laboratory (ANL), Karlsruhe Instutute of Technology and Hiroshima University to determine yields and viable detectors prior to Tracker Segment integration. 
More details on the AstroPix chip design and performance can be found in Ref.~\citeonline{Striebig_2024}.

\begin{table}[t]
    \centering{
    \begin{tabular}{|c|c|c|c|} \hline  
    Parameter & ComPair & ComPair-2 & AMEGO-X \\ \hline \hline
    \# of detectors &10   & 950 & 15200 \\  \hline
    Detector tech. & DSSD & AstroPix & AstroPix \\  \hline
    Energy res. ($\sigma$)   & 14 keV & 5 keV & 5 keV  \\ \hline
    Position res. & 500~$\mu$m \ & 500~$\mu$m & 500~$\mu$m \\ \hline
    Time res. & 10~$\mu$s \ & 1~$\mu$s & 1~$\mu$s \\ \hline
    Dynamic range & 50-700 keV & 25-700 keV & 25-700 keV \\ \hline
    \end{tabular}}
    \vspace{0.3cm}
    \caption{ \textit{The ComPair-2 Tracker performance is based on AMEGO-X requirements.  The energy resolution is reported at 122 keV. Here, we also compare with the original ComPair Tracker that was flown in 2023 and used double-sided silicon strip detectors (DSSDs).}}
    \label{tab:tracker_performance}
\end{table}

The ComPair-2 Tracker will be ten identical stacked Segments of AstroPix detectors with 5~keV energy resolution and 500~$\mu$m position resolution. 
Each Tracker Segment contains 95 Quad Chips 
that consist of four identical AstroPix APSs cut from a single wafer, as shown in \textbf{Figure~\ref{fig:compair_tracker_layer}}. 
Each Tracker Segment has a front-end electronics (FEE) board with a Lattice CertusPro-NX FPGA that configures, controls, and interfaces with the 95 Quad Chips. 
The FPGA provides the 2.5 MHz reference clock for the APS timing determination, 
and 1.2~V and 1.8~V digital and analog supplies. 
The APS I/O and power lines are daisy chained with wire bonds connecting each Quad Chip to the FEE along a thin kapton common bus bar  ($\sim$9~mm wide), as is standard practice for large arrays in ground-based silicon detectors.

The ComPair Tracker Segments' mechanical structure (see Section~\ref{sec:mechanical}) and FEE are currently being designed and fabricated at GSFC. They will be integrated at ANL 
with an automated pick-and-place machine and in-house wirebonder at the Argonne Micro Assembly Laboratory.
Fully integrated Tracker Segments will be transported back to GSFC for the full Tracker module integration and performance tests. 

In comparison to the first ComPair Tracker that was flown in 2023~\cite{kirschner2024, smith2024}, the ComPair-2 AstroPix Tracker will have an order of magnitude more detectors, a factor of three improvement in the energy resolution, and is being designed to achieve the required AMEGO-X performance as shown in \textbf{Table~\ref{tab:tracker_performance}}.

\subsection{CsI Calorimeter}

The main functionality of the Calorimeter is to measure the position and energy of Compton-scattered photons and the electromagnetic showers produced from electron and positron pairs over a broad energy range. 
The design is based on Fermi-LAT~\cite{2009ApJ...697.1071A}. Situated directly below the Tracker subsystem, the Calorimeter is composed of four layers of thallium-doped cesium iodide (CsI:Tl) bars, hodoscopically arranged.

\begin{figure}[b]
    \centering
    \includegraphics[width=0.9\textwidth, clip=true]{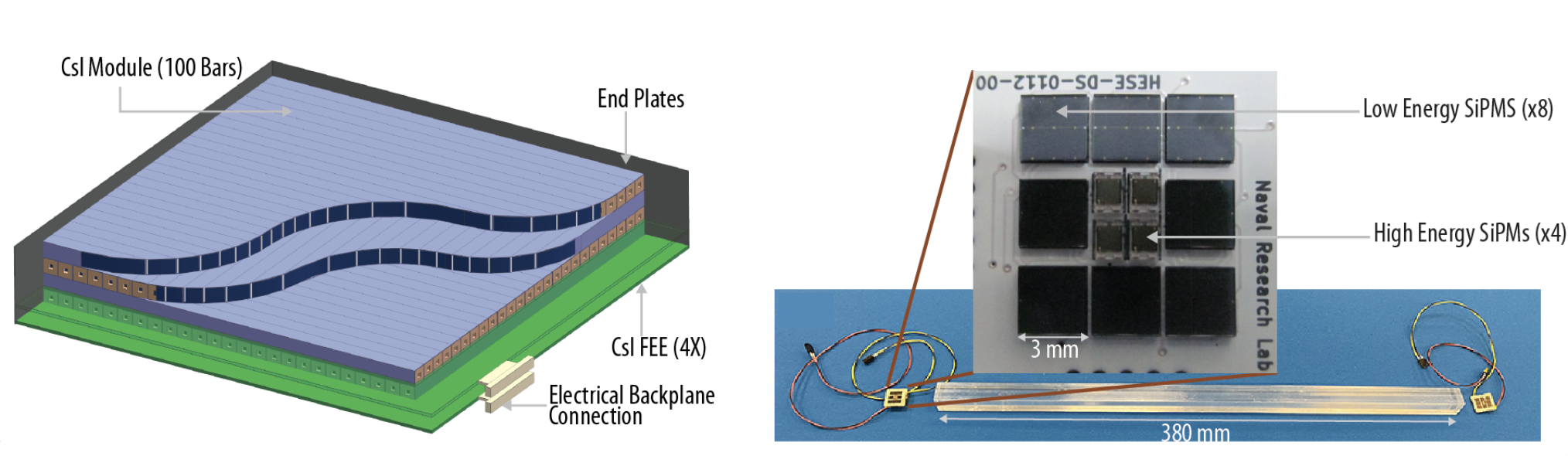}    
    \vspace{0.3cm}
    \caption{\textit{The ComPair-2 hodoscopic arrangement of four Calorimeter layers is identical to the AMEGO-X design, as shown on the left.
    (\textbf{Right}) The dual-gain protype SiPM readout implements ONSemi devices has been demonstrated through the work done at NRL. A custom design is being produced by FBK and will be used in  ComPair-2. }
    \label{fig:calorimeter}}
\end{figure}

The CsI Calorimeter for ComPair-2 will be designed, built, and tested by the Naval Research Laboratory (NRL)~\cite{CsIProceedings2023}.
Each Calorimeter layer consists of 24 CsI bars with dimensions of 1.5$\times$1.5$\times$38.8 cm$^3$ (\textbf{Figure~\ref{fig:calorimeter}}).
The bars are wrapped in reflective material to pipe scintillation photons to each end, where readout occurs via an array of SiPMs where the voltage amplitude is proportional to the energy deposited in each bar. 
The position of the interaction along the bar is determined from the relative amplitude of signals on each end.
To achieve a large dynamic range in a single Calorimeter bar and to mitigate the effects of saturation, a mixture of small (1~mm$^2$) and large (9~mm$^2$) SiPMs are used. \textbf{Figure~\ref{fig:calorimeter} \textit{right}} is a prototype of the dual gain concept implementing ONSemi SiPMs to cover two gain ranges~\cite{SiPMproceedings2023}. 
The testing campaign of this prototype dual gain array informed the design of a custom device produced by Fondazione Bruno Kessler (FBK) and intended for ComPair-2.

Each layer of the ComPair-2 Calorimeter has an L-shaped FEE board to interface with the 40 CsI logs in each layer.
The two gain channels are read out independently through the IDEAS Silicon Photomultiplier Readout ASIC (SIPHRA), which has been demonstrated with the first ComPair prototype~\cite{CsIProceedings2023}.
The SIPHRA has 16 input channels with pulse height spectroscopy, timing, and a track-and-hold with programmable hold time of up to 4.7~$\mu$s. 
The pulse height is digitized with a 12-bit analog-to-digital converter (ADC) for all channels at a sampling rate of 50 kHz.
The FEE houses six SIPHRAs per layer, which are read out and controlled with an FPGA.

\begin{table}[t]
    \centering
    \begin{tabular}{|c|c|c|c|} \hline  
    Parameter              & ComPair           & ComPair-2           & AMEGO-X \\\hline \hline
    \# of bars             & 30                     & 96                     & 384 \\  \hline
    Dim. of bars [cm]          & $1.7\times1.7\times10$ & $1.5\times1.5\times38.8$ & $1.5\times1.5\times38.8$\\\hline
    Energy res. ($\sigma$) &     20 keV             & 20 keV                 & 20 keV  \\ \hline
    Depth res. ($\sigma$)  &        2 cm            & 2 cm                   & 2 cm \\ \hline
    Dynamic range          & 0.25 - 30 MeV          & 0.1 - 400 MeV           & 0.1 - 400 MeV \\ \hline
    \end{tabular}    
    \vspace{0.3cm}
    \caption{ \textit{The ComPair-2 Calorimeter CsI bar performance is based on  AMEGO-X requirements. Energy resolution is at 662 keV. We also compare with the original ComPair Calorimeter that was flown in 2023 but did not have dual-gain SiPM readout.}}
    \label{tab:csi_performance}
\end{table}

In comparison to the first ComPair Calorimeter that was flown in 2023~\cite{shy2024}, the ComPair-2 dual-gain readout Calorimeter will have an order of magnitude larger dynamic range, with three times as many bars that each have four times the length. The ComPair-2 Calorimeter is being designed to achieve the required AMEGO-X performance as shown in \textbf{Table~\ref{tab:csi_performance}}.

\subsection{Mechanical and Thermal}
\label{sec:mechanical}
The sensitivity of MeV telescopes has historically been  limited by activation backgrounds from the instrument and spacecraft after cosmic-ray interactions~\cite{2004NewAR..48..193S}.
Therefore, the AMEGO-X Gamma-Ray Telescope design and the ComPair-2 structure implements low atomic number (Z) carbon fiber reinforced polymers to reduce backgrounds from activation and photon attenuation. 
The Tracker Segment frames and Calorimeter Module unibody structure are both fabricated and cured from M55J/Cyanate Ester (CE). 
This is a change from the original AMEGO-X design which baselined  K1100 (Thornel) fibers because of a lack of availability. 

\begin{figure}[t]
    \centering
    \includegraphics[width=0.59\textwidth]{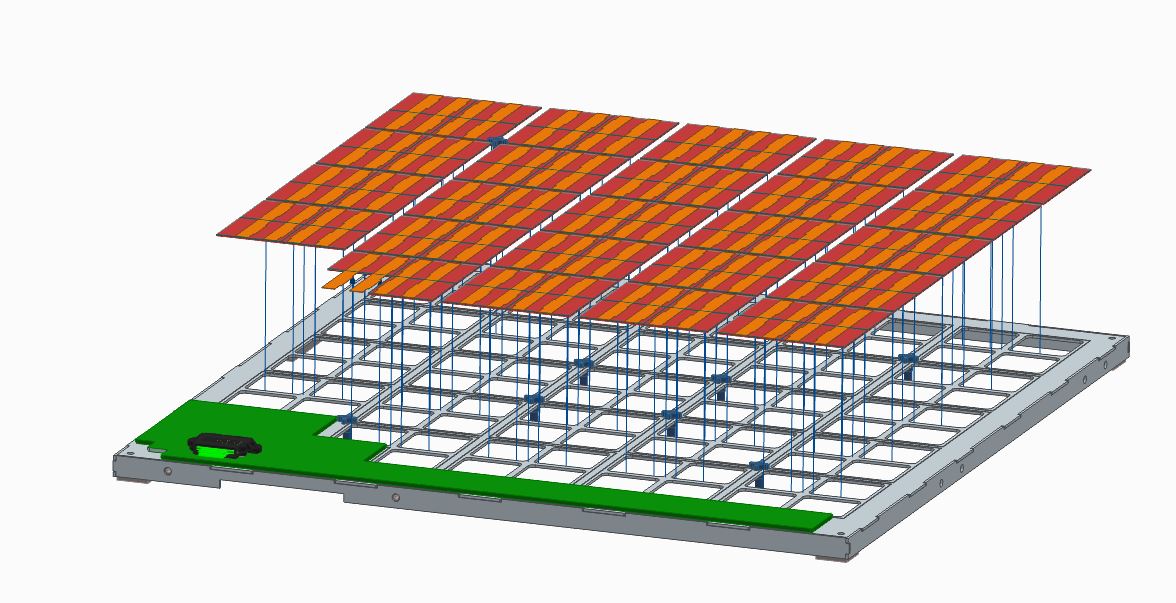}   
    \includegraphics[width=0.4\textwidth]{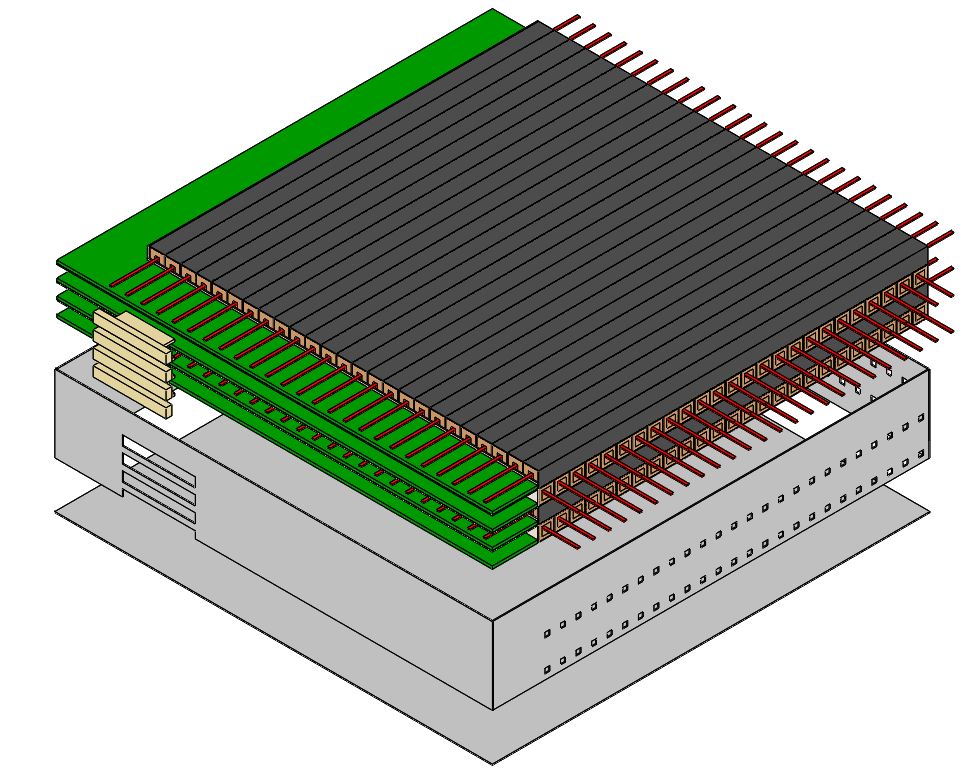} 
    \vspace{0.3cm}
    \caption{\textit{The mechanical structure for ComPair-2 is predominately M55J/CE composite to reduce the background rates from activation and limit photon attenuation. \textbf{(Left)} The Quad Chips are bonded to the Tracker Segment support frame, shown in gray, along a 2~mm periphery around window cutouts, minimizing the amount of passive material near the detectors. \textbf{(Right)} The Calorimeter has a unibody design, shown in dark gray, where each of the CsI bars slot inside and are held in place with elastomeric bumpers. The SiPM signal lines are represented as red wires here.}
    \label{fig:mechanical}}
\end{figure}

The AstroPix Tracker support frame, shown in gray in \textbf{Figure~\ref{fig:mechanical} \textit{left}}, is 0.06'' (0.15 cm) thick M55J laminate and designed with directional fibers for optimal heat extraction from the AstroPix detectors. Additional stiffness and strength is achieved with bonded M55J/CE perimeter closeouts and G10 stand-offs. 
One of the goals of the ComPair-2 project is to reduce this risk by developing automated procedures for the carbon fiber placement to produce the support frame. 
With 160 tracker segment support frames in the AMEGO-X design, to make composite fabrication using automated fiber placement can save three person years of manufacturing time compared to hand layup common to M55J production~\cite{fiberplacement2019}.
The first prototype of this mechanical frame is currently being fabricated at GSFC and will undergo vibration testing and thermal vacuum (TVAC) by the end of the year. 
The 10 layers are joined with pre-loaded tension rods and the full Tracker module has been analyzed showing minimum fundamental frequency $>$50~Hz.

The Calorimeter Module is a unibody structure, shown in dark gray in \textbf{Figure~\ref{fig:mechanical} \textit{right}}, fabricated and cured from M55J/CE based on Fermi-LAT heritage. 
Each module supports four Calorimeter layers, with alternating rows of orthogonal openings for individual CsI bars to nest inside. 
The CsI bars use an elastomeric bumper on one end for structural support and are held in place with M55J/CE End Plates, shown in light gray. 
The Calorimeter mechanical design is currently undergoing significant revisions prior to fabrication.

To extract the heat from the Tracker Segments and the Calorimeter, the AMEGO-X thermal design is based on aluminum heatpipes which have high TRL but are not conducive to flexible bench-top operations. 
Instead, the ComPair-2 thermal design will use a liquid cooling loop based on the heritage balloon thermal extraction systems that will operate in the lab and in thermal vacuum testing. 
By having the same connection profile to the Tracker Segments and a similar efficiency to heat pipes, the cooling loop solution validates the thermal design at the individual layer level. 
An aluminum thermal plate that is the same profile as the AMEGO-X heatpipe will be thermally coupled to each Tracker and Calorimeter layer.  Heat will be extracted via a liquid cooling loop connected to a large radiator with a cold sink. 

\subsection{Data Acquisition and Triggering}
\label{sec:daq}

\begin{figure}
    \centering
    \includegraphics[width=0.40\textwidth, clip=true]{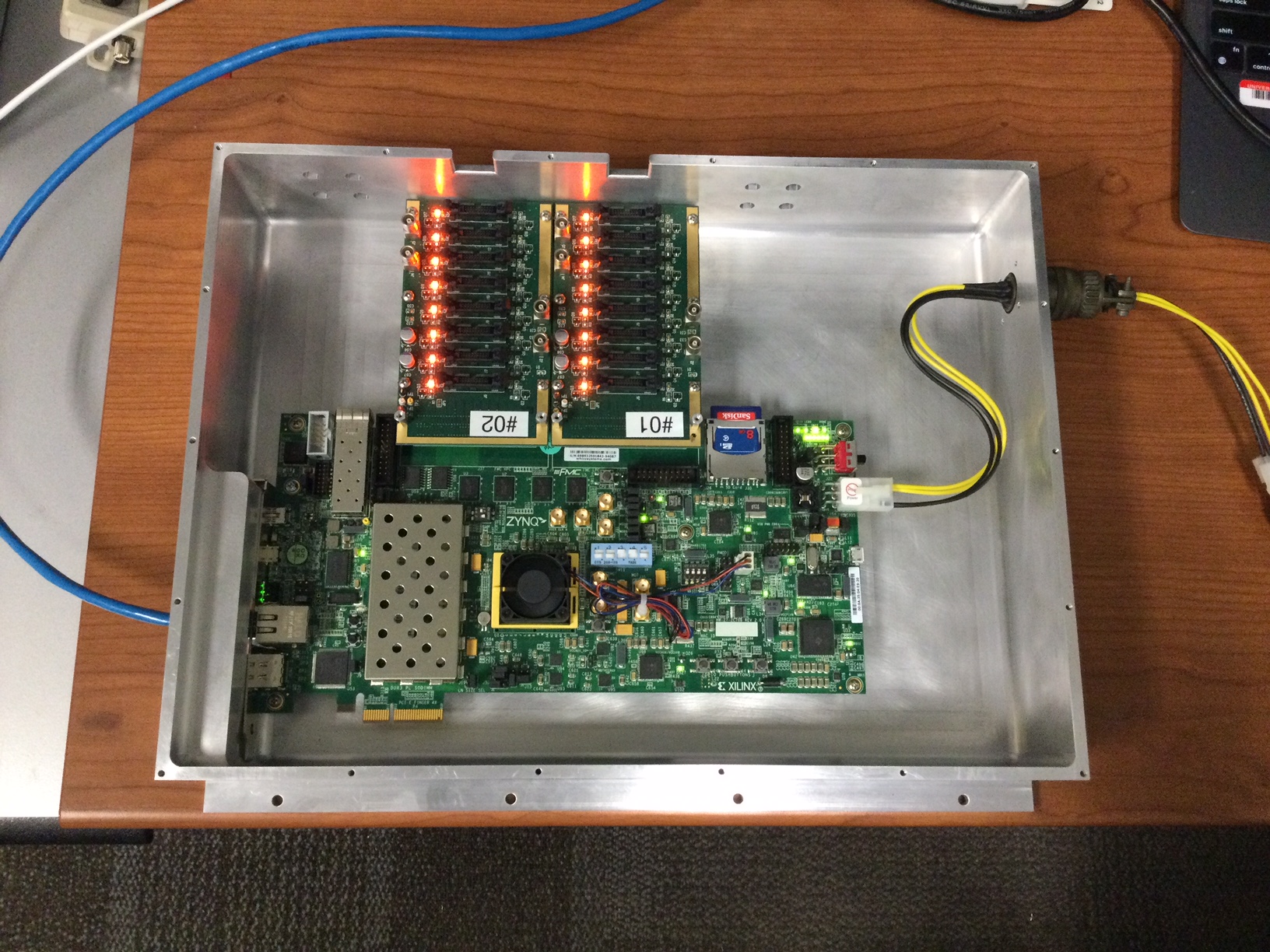}  
    \vspace{0.3cm}
    \caption{\textit{The ComPair-2 Trigger Module will be reused from the ComPair balloon payload. It implements a Xilinx ZC706 FPGA evaluation board (shown in its Aluminium housing) connected to a custom interface board, shown with LEDs on at the top of the box. }
    \label{fig:trigger_module}}
\end{figure}

The ComPair-2 DAQ will be reused from the original ComPair balloon payload, which has a Trigger Module (TM) for the off-line combination of coincidence events~\cite{sasakispie} 
and a flight CPU for data readout. 
The TM (\textbf{Figure~\ref{fig:trigger_module}}) uses a COTS Xilinx ZC706 FPGA evaluation board.
Each subsystem segment sends fast ($1~\mu$s) trigger primitives to the TM which checks for coincidence within a programmable time window. 
Only events with more than one interaction in the Tracker, or a coincident detection in the Tracker and Calorimeter are considered ``good events.''
In addition to reducing the telemetered data rate during a future AMEGO-X mission, this trigger logic also serves to reduce the deadtime of the instrument by initiating the slow readout of the Calorimeter bars ($\sim$6 $\mu$s) only when the logic is satisfied. 
The TM has 16 prescalable channels that allow for sophisticated trigger combinations and a low-level monitoring of raw rates. 

In addition to the basic readout and control, this proposed work supports the development and validation of real-time event reconstruction algorithms in the CPU. 
This novel effort would allow a faster localization ($<$30 s vs. $>$30 mins) of transients by performing on-board simplified Compton event reconstruction.
The LANL team has performed a proof-of-concept for this approach with simulated idealized data on a high-powered FPGA. 
The fully integrated ComPair-2 with a flight-like CPU will provide the first hardware demonstration of on-board Compton localization in real time.

%% file: IandT.tex
With over 15000 detectors (see \textbf{Table~\ref{tab:tracker_performance}}) the integration and testing process for AMEGO-X is a large driver of the schedule and cost of the mission. 
One of the goals of the ComPair-2 project is to reduce this risk by developing automated procedures the integration and testing of each Tracker Segment. 
There are two prototype Segments frames currently being fabricated that will be sent to ANL for integration with dummy silicon wafers and AstroPix v3 chips. 
These prototype segments will undergo vibration testing and TVAC testing to confirm the mechanical and thermal performance meets the requirements, and the integration procedure is well defined. After these two Segments are validated, the final 10 ComPair-2 Tracker Segments will be fabricated and integrated.

The Tracker and Calorimeter modules will be individually integrated by the end of 2025, and 2026 will be dedicated to the integration and test of the full ComPair-2 instrument.
Testing will be comprised of four different campaigns: low-energy ($<$1~MeV) calibrations and performance validation in the Compton regime with laboratory radioactive sources; thermal vacuum testing at GSFC; vibration tests at GSFC; and high-energy ($>$1~MeV) performance validation at the polarized High Intensity Gamma-ray Source (HIGS). 
All of these efforts, have an established process developed through the ComPair balloon payload~\cite{2022SPIE12181E..2GS}.

These tests will provide accurate measurements of the instrument and subsystem detection efficiency, energy and angular resolution, and polarization sensitivity in a flight-like prototype. 
While this testing will primarily raise the Technical Readiness Level of the AMEGO-X Gamma-Ray Telescope, this effort will also prepare for a future ComPair-2 balloon flight. 

%% file: Pipelines.tex
\begin{figure}
    \centering
    \includegraphics[width=0.6\textwidth]{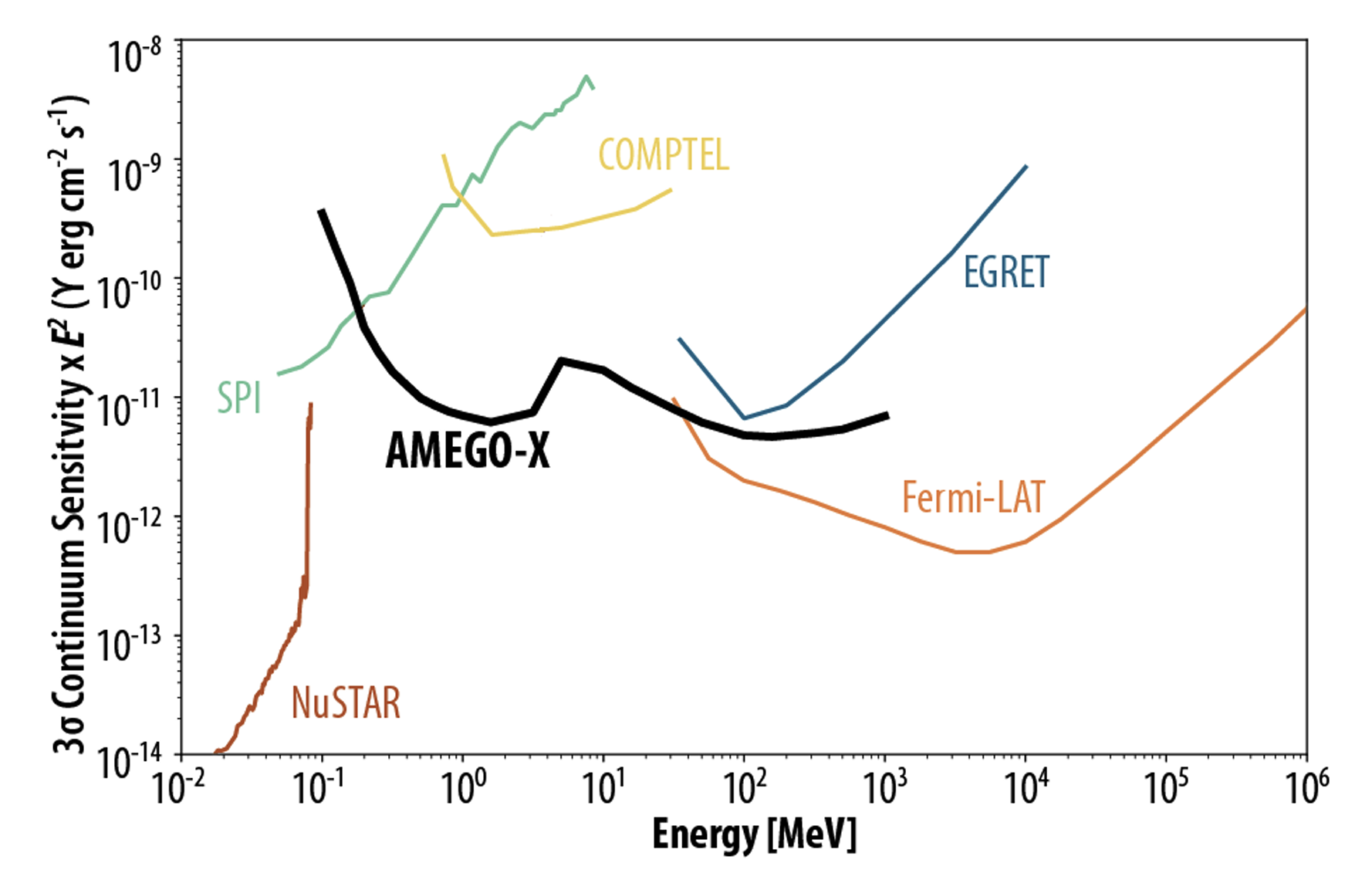}
    \includegraphics[width=0.39\textwidth, clip=true]{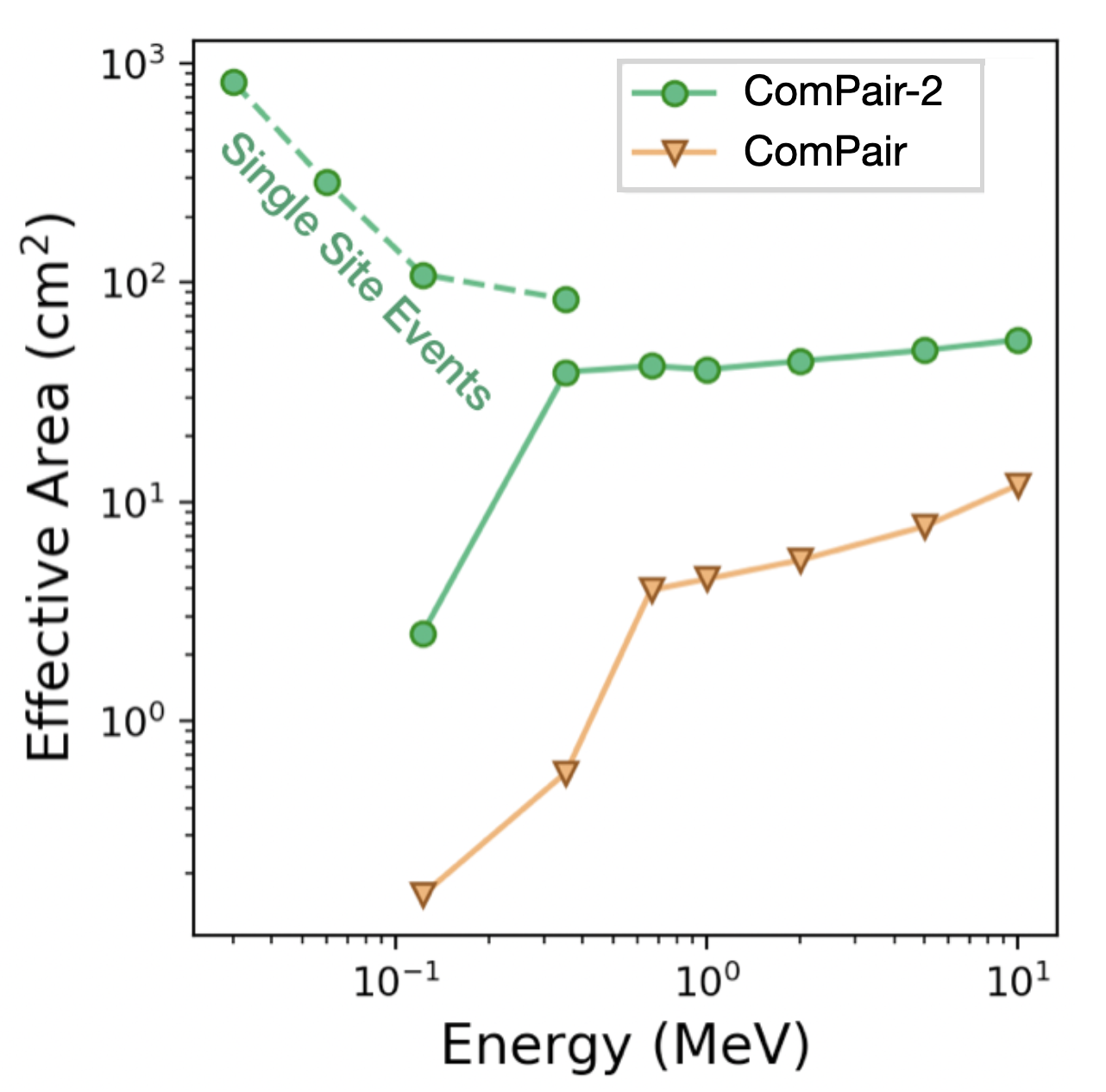}
    \caption{\textit{(\textbf{left}) The AMEGO-X sensitivity curve using standard reconstruction algorithms available in MEGAlib. The characteristic ``W'' shape comes from challenges in Compton and pair identification in the cross-over region around 10~MeV. This can be addressed with better reconstruction algorithms. (\textbf{right}) ComPair-2 will achieve an order of magnitude increase in effective area compared to the first ComPair balloon payload, even before pipeline optimizations. Furthermore, the ComPair-2 team will implement 3D convolutional neural network algorithms to correct the challenges in the Compton and pair classification.}
    \label{fig:compair_performance}}
\end{figure}

All ComPair analysis uses the Medium Energy Gamma-ray Astronomy library (MEGAlib), which is the state-of-the-art software tool for MeV telescopes~\cite{MEGALIB}. 
MEGAlib performs Geant4-based Monte Carlo simulations, event identification and reconstruction, and high-level analysis. The AMEGO-X performance and sensitivity calculations, shown in \textbf{Figure~\ref{fig:compair_performance} \textit{left}}, have been performed in MEGAlib and are described in Ref.~\citeonline{AMEGOX}. 

The event reconstruction for Compton and pair telescopes is relatively complicated and can be the limiting factor for the sensitivity of MeV instruments. 
For example, the ``W'' shape in the AMEGO-X sensitivity curve comes from inaccurate Compton and pair identification in the cross-over region around 10~MeV (\textbf{Figure~\ref{fig:compair_performance} \textit{left}}). 
ComPair's Berkeley/SSL team has shown that a 3D convolutional neural network algorithm achieves over 99\% correct identification for simulated Compton and pair events in an AMEGO-X like instrument~\cite{zoglauer2021}. 
This approach outperforms the accuracy of event identification used by the current 
ComPair balloon instrument by almost a factor of two, which directly leads to an increased effective area by the same fraction. 
Furthermore, this work explored machine learning techniques for transient localization, Compton electron tracking reconstruction, and pair event energy estimation. 
While these are all encouraging results for the next-generation of MeV telescopes, much of this work has not yet been demonstrated with laboratory data.

The ComPair-2 team is working to implement novel event reconstruction techniques for the Compton and pair regime into MEGAlib. 
These tools will be validated with ComPair-2's laboratory and gamma-ray beam data. 
Ultimately, this work will enable more accurate sensitivity predictions at the particularly challenging $\sim$10 MeV energy, and is expected to flatten the ``W'' shape in the sensitivity curve. 

Even without these improvements in the pipeline, there is still significant improvement in the ComPair-2 instrument compared to ComPair. \textbf{Figure~\ref{fig:compair_performance} \textit{right}} 
shows the combined Compton and pair effective area of both instruments as a function of energy, highlighting the order of magnitude increase before the proposed software optimization. This improvement is due to the incorporation of AstroPix detectors (lower thresholds and improved noise performance), and the larger sensitive area of the instrument.
Single-site events in the Tracker enable observations of transients at low energies (see \textbf{Section~\ref{sec:conops}}).

%% file: Conclusions.tex
ComPair-2 is a prototype of a next-generation gamma-ray telescope.
The goals of this prototype are threefold:
incorporate the AstroPix detectors and increase the effective area of the Tracker;
confirm telescope performance across the Compton and pair regimes and in a relevant environment, raising the TRL to 6; and
provide the first hardware demonstration of novel event reconstruction techniques.
After the successful completion of ComPair-2, we will be ready to fly on a long-duration balloon flight with follow-on support. 
Furthermore, this effort builds off of the AMEGO-X MIDEX design, and the successful completion will raise the technical readiness level (TRL) of the AMEGO-X Gamma-Ray Telescope from 4 to 6, directly leading to a more mature submission for the next MIDEX opportunity.
In addition to TRL advancement, this proposed research provides valuable experience and necessary hands-on training to the early-career ComPair management, postdocs, graduate students, and engineers within the diverse team. 